\documentclass{article}

\usepackage{arxiv}

\usepackage{lipsum}

\usepackage[utf8]{inputenc} 
\usepackage[T1]{fontenc}    
\usepackage{hyperref}       
\usepackage{url}            
\usepackage{booktabs}       
\usepackage{amsfonts}       
\usepackage{nicefrac}       
\usepackage{microtype}      
\usepackage[ruled,vlined]{algorithm2e}

\usepackage{multirow}

\usepackage{caption}
\usepackage{subcaption}

\usepackage[utf8]{inputenc} 
\usepackage[T1]{fontenc}    
\usepackage{hyperref}       
\usepackage{url}            
\usepackage{booktabs}       
\usepackage{amsfonts}       
\usepackage{nicefrac}       
\usepackage{microtype}      
\usepackage{amsmath}
\usepackage{amsthm}
\usepackage{enumitem}

\usepackage{tikz}
\usetikzlibrary{arrows.meta}
\tikzset{%
  >={Latex[width=2mm,length=2mm]},
            base/.style = {rectangle, rounded corners, draw=black,
                           minimum width=4cm, minimum height=1cm,
                           text centered, font=\sffamily},
  activityStarts/.style = {base},
       startstop/.style = {base},
    activityRuns/.style = {base},
         process/.style = {base, minimum width=2.5cm,
                           font=\sffamily},
}

\setitemize{topsep=0pt,parsep=0pt,partopsep=0pt,leftmargin=*}

\usepackage{thmtools}
\usepackage{thm-restate}

\usepackage{xcolor}

\definecolor{vega10-0}{HTML}{1f77b4}
\definecolor{vega10-1}{HTML}{ff7f0e}
\definecolor{vega10-2}{HTML}{2ca02c}
\definecolor{vega10-3}{HTML}{d62728}
\definecolor{vega10-4}{HTML}{9467bd}
\definecolor{vega10-5}{HTML}{8c564b}
\definecolor{vega10-6}{HTML}{e377c2}
\definecolor{vega10-7}{HTML}{7f7f7f}
\definecolor{vega10-8}{HTML}{bcbd22}
\definecolor{vega10-9}{HTML}{17becf}

\usepackage{tikz}
\usetikzlibrary{bayesnet}
\usetikzlibrary{calc}

\title{Accelerometer-Based Gait Segmentation: Simultaneously User and Adversary Identification} 

\author{%
  Yujia Ding
  \\
  Institute of mathematical sciences\\
  Claremont Graduate University\\
  Claremont, CA 91711 \\
  \texttt{yujia.ding@cgu.edu} \\
   \And
   Weiqing Gu \\
   Department of Mathematics \\
   Harvey Mudd College\\
   Claremont, CA 91711 \\
   \texttt{gu@g.hmc.edu} \\
}

\begin{document}

\maketitle

\begin{abstract}
In this paper, we introduce a new gait segmentation method based on accelerometer data and develop a new distance function between two time series, showing novel and effectiveness in simultaneously identifying user and adversary. Comparing with the normally used Neural Network methods, our approaches use geometric features to extract walking cycles more precisely and employ a new similarity metric to conduct user-adversary identification. This new technology for simultaneously identify user and adversary contributes to cybersecurity beyond user-only identification. In particular, the new technology is being applied to cell phone recorded walking data and performs an accuracy of $98.79\%$ for 6 classes classification (user-adversary identification) and $99.06\%$  for binary classification (user only identification). In addition to walking signal, our approach works on walking up, walking down and mixed walking signals. This technology is feasible for both large and small data set, overcoming the current challenges facing to Neural Networks such as tuning large number of hyper-parameters for large data sets and lacking of training data for small data sets.  In addition, the new distance function developed here can be applied in any signal analysis.     
  
\end{abstract}
\section{Introduction}
\label{introduction}
In today’s technology-advanced and data-explosion world, cell phone usage is woven into nearly every aspect of human life. As we enjoy the major benefits provided by cell phones, the rapid growing use of the information network brings with it a rise in malicious cyber activities, and hence proves strong reasons for cybersecurity. Evidences have shown that approaches based on cell phones data that contains both accelerometer and gyroscope signals usually provides good performance \cite{neverova2016learning}. However, there are a fair number of cell phones have no  gyroscope data recorded, especially for some low-priced cell phones. This paper proposes new techniques for gait extraction on accelerometer signal only; identification performance can be improved by adding gyroscope data when it is available. Moreover, our gait analysis is applied on human motion identification for strong user authentication on smart phones. The current authentication process for smart phones is largely based on weak static credentials, such as passwords or swipe patterns on touchscreens. Such credentials usually has many vulnerabilities that the attacker can easily gain access to the cell phone. Therefore, this paper researches on authentication methods aim to constantly protect the cell phone and simultaneously detecting the adversary. In addition to advanced knowledge on improving cell phone security, the methods provided in this paper will have several broader impacts; the gait segmentation approach works for other human motion analysis and the new time series distance, a new way of measuring similarity between signals, can be applied in other signal analysis. The principal arguments underlying this paper including: (a) a well-performed data pre-processing by the gait segmentation method that wipes off the non-walking activities and accurately segments the walking series into gait cycles is discussed in Section \ref{segmentation_algorithm}; (b) a new measurement for signal distance that transforms human vision into machine language is proposed in Section \ref{signal_distance}; (c) experimental studies of segmentation as well as a fast, stable and accurate  user-adversary identification are demonstrated in Section \ref{experimental}; (d) a summary and conclusion of the impact of this research is shown in Section \ref{discussion}.

Gait cycle segmentation or classification is broadly pre-processing step in studying human walking. In particular, high accuracy of gait cycle segmentation is extremely important in increasing the performance of human authentication. Another crucial usage of gait segmentation is to monitor or diagnose specific diseases such as Parkinson’s disease or peripheral neuropathy. There are many systems can be used to collect gait signals, such as vision-based systems, goniometers, Inertial Measurement Units (IMUs) and foot pressure sensors, among which IMUs is mostly studied in this paper. 
Many gait cycle segmentation methods have been studied based on different types of signal recorded from different types of sensors. Some gait segmentation methods are applied in identification of contact events,
such as heel strike and toe-off which could
be particularly useful in providing online assistance during
walking; different type of signals have been used, such as
only kinematics data from the knee and hip joints \cite{kalinowska2019data} or angular velocity of lower limb \cite{grimmer2019stance}.  Some Hidden Markov Model based machine-learning methods are investigated in gait segmentation by pressure sensors, see \cite{crea2012development, de2012gait}. Foot-switch signal gait segmentation is researched in \cite{agostini2013segmentation}. Comparison of four gait segmentation methods, including peak detection from event-based methods, two variations of dynamic time warping from template matching methods, and hierarchical hidden Markov models from machine learning methods, are made to exam the Parkinsonian gait in \cite{haji2018segmentation}. Researches on pathological gait abnormality detection and
segmentation are made in \cite{khan2019pathological}. In the literature, zero-crossing and  Lomb-Scargle periodogram for cycle extraction are widely used; the former is based on wave form \cite{khan2019pathological, sugandhi2019overlap} and the latter is for detecting and characterizing periodic signals in unevenly-sampled data \cite{vanderplas2018understanding}. A robust algorithm for gait cycle
segmentation is provided based on a peak detection approach in \cite{jiang2017robust}. Many accelerometer-based gait analysis has been summarized in \cite{D2010}. 

\begin{figure}[h]
\centering
\begin{subfigure}{0.4\columnwidth}
\centering
\includegraphics[width=\textwidth]{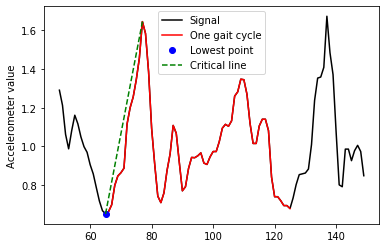}
\caption{}
\end{subfigure}
\begin{subfigure}{0.4\columnwidth}
\centering
\includegraphics[width=\textwidth]{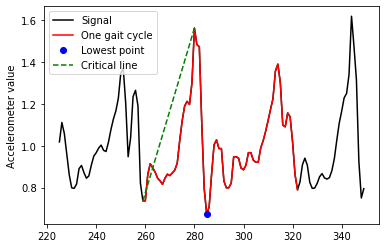}
\caption{}
\end{subfigure}
\caption{Gait cycle on x-axis of accelerometer data for walking. Left (a): data from HAPT volunteer 30; Gait cycle starts with point (65, 0.65) and ends with point (125, 0.68). Right (b): data from HAPT volunteer 3; Gait cycle starts with point (259, 0.74) and ends with point (323, 0.79); the lowest value in the cycle is at point (285, 0.67).}
\label{gait_cycle}
\end{figure}

Existing gait segmentation algorithms have achieved outstanding performance, but showed its deficiency when dealing with very noisy data and activity mixed gait signals. Therefore, the accelerometer-based segmentation method proposed in this paper aims to address the above problems and improves the performance of user-adversary identification or, in particular, the accuracy of smart phone authentication. The gait cycle obtained by our algorithm began when one foot touched the ground and ended when the same foot touched the ground again, shown in Figure \ref{gait_cycle} with x-axis of the accelerometer data. Several geometry features of the accelerometer-based gait cycle are summarized in the following:
\begin{enumerate}
    \item[$\mathcal{F}_1$:] Both of the starting and ending  points of the gait cycle are minimal peaks with bigger angles defined in Equation (\ref{angle_def}).
    \item[$\mathcal{F}_2$:] Values of the start or end point of the gait cycle are lower than most of the points in the series and are the lowest within a cycle in most cases. See the left plot in Figure \ref{gait_cycle} for an example of common gait cycles and an exception is given in the right plot in Figure \ref{gait_cycle}.
    \item[$\mathcal{F}_3$:] The start and end points of the cycle are on critical or dominated lines. The critical line is the longest vector in the cycle that ignores small peaks as the dash lines indicated in Figure \ref{gait_cycle}.
\end{enumerate}
The robust algorithm for gait cycle
segmentation provided in \cite{jiang2017robust}, however, can not deal with the cases when in $\mathcal{F}_2$, the values of the edge points are not the lowest within a cycle, see for example the right plot in Figure \ref{gait_cycle}. Our proposed gait segmentation approach is more reliable when dealing with real walking signal that involves all but not limited to the above problems and results in more accurate segmentation.\\

Apparently, it becomes confused when we mention the accuracy of the segmentation. We can definitely visually judge if the segmentation satisfies our criteria. However, teaching computer to determine if a cutting is good or not under the same criteria is a difficult task; many misjudgement are really out there and better approaches are urgently needed. We propose a simple and fast signal distance measurement that plays the role of translating human vision to computer language. This method provides the score of tuning the three hyper-parameters involved in the segmentation and as a result decides the accuracy of the segmentation. The segmentation technology as well as the signal distance investigated in this paper should not be limited to be applied in the user-adversary identification problem. Many applications in the area of human motion analysis and other type of signal analysis requires high accuracy method described here to address real world problems.  

The vision-based human motion analysis in \cite{zeng2019vision} automatically identify the behaviour of human from a given image or a sequence of images. A human motion recognition method based on Kalman random forest algorithm model is studied in \cite{yi2019human} for improving the accuracy and efficiency of tracking algorithm. Many neural-network based methods to recognize human motion intention are widely investigated in literature, see e.g. \cite{wu2019neural, lang2019joint, neverova2016learning, oukrich2019daily, furuya2019personal}.  A method for positive identification of smartphone user’s identity is presented in \cite{damavsevivcius2016smartphone}, using user’s gait characteristics based on the application of the Random Projections method for feature dimensionality reduction.
 
Relying on the segmentation approach as well as the signal distance proposed, the accuracy of our user-adversary identification are rapidly improved even with simple and native classification methods. Against the current existing methods, we develop a new distance function for our classification method based on novel techniques of extracting certain archetypes to identify user and adversary behaviors from the training data set. Specifically, we extract the archetypes that represent individual walking behavior by simple clustering methods based on our signal distance. Then comparison are made between new signal data to the known archetypes of candidates containing user and adversary. The overall accuracy obtained shows that our methods are well-performed in cell phone authentication and detecting theft. 

\section{Accelerometer-based segmentation algorithm}
\label{segmentation_algorithm}

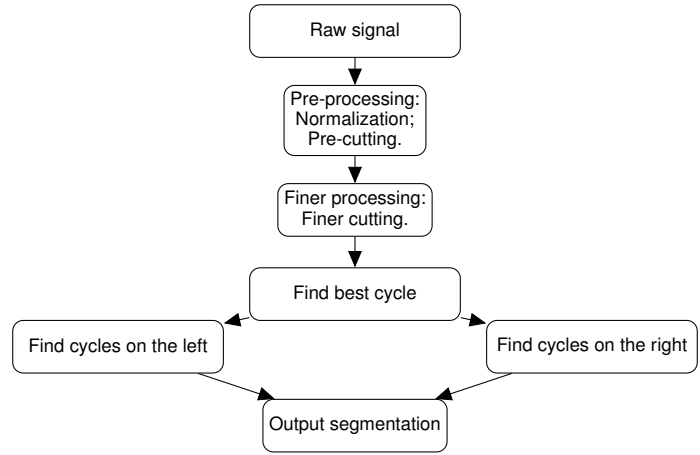
\begin{figure}[!htb]
\begin{center}
\begin{tikzpicture}[node distance=1.5cm,
    every node/.style={fill=white, scale=0.7, font=\sffamily}, align=center]
    \label{workflow}
  \node (start)             [activityStarts]              {Raw signal};
  \node (onStartBlock)      [process, below of=start, yshift=-0.2cm]   {Pre-processing:\\
  Normalization;\\
  Pre-cutting.};
  \node (onResumeBlock)     [process, below of=onStartBlock, yshift=-0.2cm]   {Finer processing:\\
       Finer cutting.
  };
  \node (activityRuns)      [activityRuns, below of=onResumeBlock, yshift=-0.1cm]    {Find best cycle};
  \node (onPauseBlock)      [process, below of=activityRuns, yshift=-1cm]
                                                                {Output segmentation};
  \node (ActivityEnds)      [startstop, left of=activityRuns, xshift=-3cm, yshift=-1cm]
                                                        {Find cycles on the left};
  \node (ActivityEnds_1)      [startstop, left of=activityRuns, xshift=6cm, yshift=-1cm]
                                                        {Find cycles on the right};                                     
\draw[->]             (start) -- (onStartBlock);
  \draw[->]      (onStartBlock) -- (onResumeBlock);
  \draw[->]     (onResumeBlock) -- (activityRuns);
  \draw[->]      (activityRuns) --  (ActivityEnds_1);
  \draw[->]      (activityRuns) --  (ActivityEnds);
  \draw[->]      (ActivityEnds) --  (onPauseBlock);
  \draw[->]      (ActivityEnds_1) --  (onPauseBlock);
  \end{tikzpicture}
  \end{center}
  \caption{Segmentation algorithm flowchart}
\label{segmentation_algorithm_flowchart}
\end{figure}

In this section, a detailed gait segmentation algorithm is introduced based on the geometry features discussed in Section \ref{introduction}. The flow chart in Figure \ref{segmentation_algorithm_flowchart} briefly introduces the process of segmentation as well as the organization of this section. 

\begin{figure}[h]
\centering
\begin{subfigure}{\columnwidth}
\centering
\includegraphics[width=0.8\textwidth]{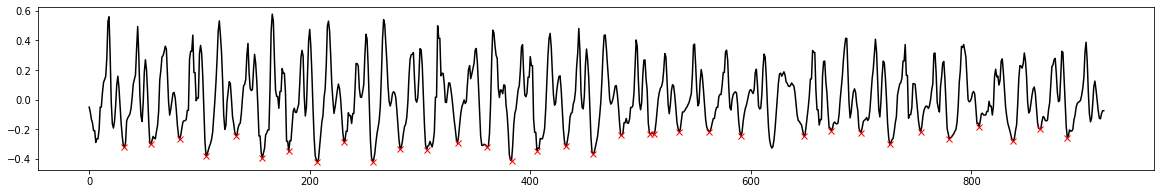}
\caption{}
\end{subfigure}
\begin{subfigure}{\columnwidth}
\centering
\includegraphics[width=0.8\textwidth]{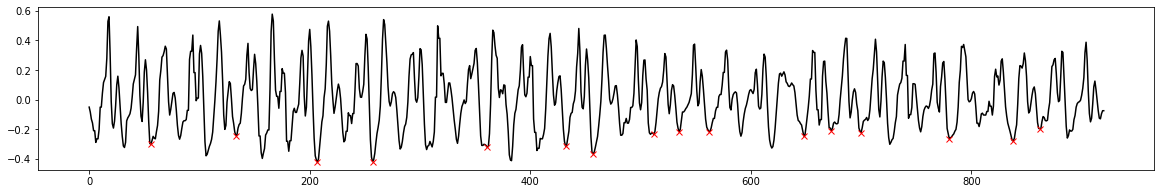}
\caption{}
\end{subfigure}
\begin{subfigure}{\columnwidth}
\centering
\includegraphics[width=0.8\textwidth]{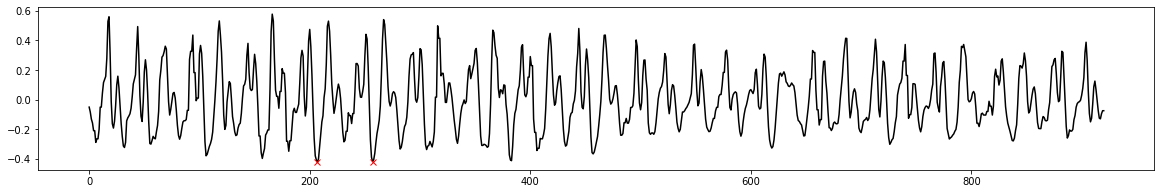}
\caption{}
\end{subfigure}
\begin{subfigure}{\columnwidth}
\centering
\includegraphics[width=0.8\textwidth]{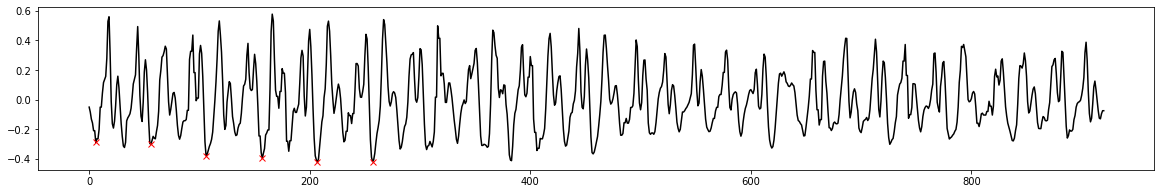}
\caption{}
\end{subfigure}
\begin{subfigure}{\columnwidth}
\centering
\includegraphics[width=0.8\textwidth]{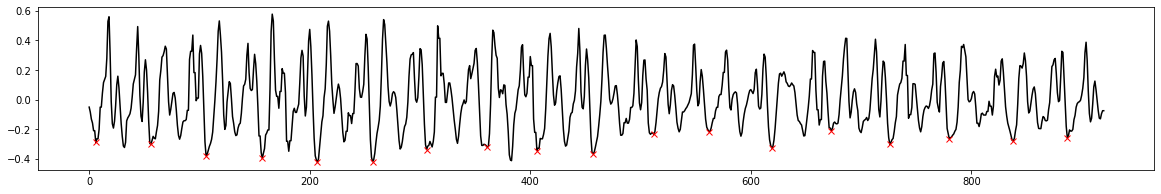}
\caption{}
\end{subfigure}
\caption{X-axis full accelerometer segmentation processes on 1st period of volunteer 9 in HAPT data set with walking signal. 1st plot (a): pre-processing (Section \ref{pre_processing}); 2nd plot (b): finer processing (Section \ref{finer_processing}); 3rd plot (c): find best cycle (Section \ref{find_best_cycle}); 4th (d) and 5th plot (e): find cycles on the left and right of the best cycle (Section \ref{find_all_cycles}).}
\label{find_cycle_example}
\end{figure}

In Subsection \ref{pre_processing}, several pre-cutting detection approaches are applied to obtain a rough segmentation, which follows Subsection \ref{finer_processing}, where finer segmentation is processed. Next, a best cycle is determined from the finer segmentation in Subsection \ref{find_best_cycle}. Finally, Subsection \ref{find_all_cycles} talks about approaches to obtain the final segmentation. For an example of segmentation processes, see Figure \ref{find_cycle_example}. In the following, let $S = (s(1),\ldots, s(N))'$ denotes the raw signal series, which has $N$ data points in total and $s(t)$ be the value of the $t$th data on the signal. Our goal is to find the segmentation of a given signal, where segmentation is represented by all the cycles on the signal, or in other words, all points that divide the signal into cycles. We call these dividing points "cuts".  Here cycle is the partial signal between two nearby cuts.

\subsection{Pre-processing} 
\label{pre_processing}
Features $\mathcal{F}_2$ and $\mathcal{F}_3$ provided an idea of identify cutting positions. Apparently, we are more interested in the points on critical lines (dominated lines) and at the same time has relatively smaller y-values than others.
We will use peak detection method provided by Scipy to roughly find all minimal peaks of the raw signal, before which, normalization is required to standardize different signals so that the peak detection method can be applied under same criteria. The minimal peaks obtained are the pre-cuts of the signal.
\begin{enumerate}
    \item First, we scale the signal $S$ to have the range of $[0,1]$. The $t$th scaled data $s^{\scriptsize\mbox{sc}}(t)$ is:
    \begin{equation}
    \label{normalize_scale}
    s^{\scriptsize\mbox{sc}}(t) = \frac{s(t)-\min(S)}{\max(S)-\min(S)}.    
    \end{equation}
    \item Second, we shift the scaled signal $S^{\scriptsize\mbox{sc}}=(s^{\scriptsize\mbox{sc}}(1),\ldots, s^{\scriptsize\mbox{sc}}(N))'$ to have $0$ mean. The $t$th $0$-mean data $s^{\scriptsize\mbox{sh}}(t)$ is:
    \begin{equation}
    \label{normalize_shift}
    s^{\scriptsize\mbox{sh}}(t) = s^{\scriptsize\mbox{sc}}(t)-\frac{1}{N}\sum_{j=1}^N s^{\scriptsize\mbox{sc}}(j).   
    \end{equation}
    \item Last, we apply the peak detection in scipy on the normalized signals $S^{\scriptsize\mbox{sh}}=(s^{\scriptsize\mbox{sh}}(1),\ldots, s^{\scriptsize\mbox{sh}}(N))'$ with the same criteria, to obtain the pre-cuts. Guaranteed by (\ref{normalize_scale}) and (\ref{normalize_shift}),  height feature involved in the peak detection method in Scipy can be set to be $[0.1,0.5]$ for all signals. Moreover, the width feature is set to be $10$ for simplicity, since eventually the cycle cutting points are expected to be not very concentrated.
\end{enumerate}
 The result of the pre-cutting process is a vector of points $(x^{\scriptsize\mbox{pr}}_i,y^{\scriptsize\mbox{pr}}_i)$ for $i=1,\ldots,n^{\scriptsize\mbox{pr}}$ with $n^{\scriptsize\mbox{pr}}\leq N.$ Here $x^{\scriptsize\mbox{pr}}_i$ denotes the position of the $i$th data in $S$ and $y^{\scriptsize\mbox{pr}}_i=s(x^{\scriptsize\mbox{pr}}_i)$ denotes the value of the $i$th data in $S$.
The mappings involved in (\ref{normalize_scale}) and (\ref{normalize_shift}) are invertible and have no change on the time axis $t$ of the signal, so the positions of the $i$th pre-cut is $x^{\scriptsize\mbox{pr}}_i$, which is the same on $S$, $S^{\scriptsize\mbox{sc}}$ and $S^{\scriptsize\mbox{sh}}$.

\subsection{Finer processing}
\label{finer_processing}
 The cuts suppose to be the points where a gait cycle starts or ends, so the angles of the peaks in the accelerometer plots suppose to be relatively bigger, which is due to $\mathcal{F}_1$ of gait cycle discussed in Section \ref{introduction}. Therefore, we pick peaks with larger angles to be possible candidates of the segmentation cuts. Recall that the $i$th pre-cut point is $\left(x^{\scriptsize\mbox{pr}}_i,y^{\scriptsize\mbox{pr}}_i\right)$, the closest points left and right to $\left(x^{\scriptsize\mbox{pr}}_i,y^{\scriptsize\mbox{pr}}_i\right)$ on $S$ are $\left(x^{\scriptsize\mbox{pr}}_i-1,s(x^{\scriptsize\mbox{pr}}_i-1)\right)$ and $\left(x^{\scriptsize\mbox{pr}}_i+1,s(x^{\scriptsize\mbox{pr}}_i-1)\right)$ respectively. The angle $\alpha_i$ of data point $\left(x^{\scriptsize\mbox{pr}}_i,y^{\scriptsize\mbox{pr}}_i\right)$ is defined as the angle between vectors $\left(x^{\scriptsize\mbox{pr}}_i-1,s(x^{\scriptsize\mbox{pr}}_i-1)\right)-\left(x^{\scriptsize\mbox{pr}}_i,y^{\scriptsize\mbox{pr}}_i\right)$ and $\left(x^{\scriptsize\mbox{pr}}_i+1,s(x^{\scriptsize\mbox{pr}}_i+1)\right)-\left(x^{\scriptsize\mbox{pr}}_i,y^{\scriptsize\mbox{pr}}_i\right)$:
\begin{equation}
\label{angle_def}
    \cos\alpha_i = \frac{\left(s(x^{\scriptsize\mbox{pr}}_i-1)-y^{\scriptsize\mbox{pr}}_i\right)\left(s(x^{\scriptsize\mbox{pr}}_i+1)-y^{\scriptsize\mbox{pr}}_i\right)-1}{\left(\left(s(x^{\scriptsize\mbox{pr}}_i-1)-y^{\scriptsize\mbox{pr}}_i\right)^2+1\right)^{1/2}\left(\left(s(x^{\scriptsize\mbox{pr}}_i+1)-y^{\scriptsize\mbox{pr}}_i\right)^2+1\right)^{1/2}}.
\end{equation}
Then we pick cuts from pre-cuts with angles above the median of the  pre-cut angles. Now the finer processed cuts are denoted as $(x^{\scriptsize\mbox{fi}}_i,y^{\scriptsize\mbox{fi}}_i)$ for $i=1,\ldots,n^{\scriptsize\mbox{fi}}$ with $n^{\scriptsize\mbox{fi}}<n^{\scriptsize\mbox{pr}}\leq N.$ Here $x^{\scriptsize\mbox{fi}}_i$ denotes the position of the $i$th point on $S$ and $y^{\scriptsize\mbox{fi}}_i=s(x^{\scriptsize\mbox{fi}}_i)$ denotes the value of the $i$th pre-cut point on $S$.
\subsection{Find best cycle}
\label{find_best_cycle}
 Based on the finer cuts $(x^{\scriptsize\mbox{fi}}_i,y^{\scriptsize\mbox{fi}}_i),$ $i=1,\ldots,n^{\scriptsize{\mbox{fi}}}$ picked in last subsection \ref{finer_processing}, cycles involved in $(x^{\scriptsize\mbox{fi}}_i,y^{\scriptsize\mbox{fi}}_i),$  $i=1,\ldots,n^{\scriptsize{\mbox{fi}}}$ are denoted by
 $$
 \mathbf C^{\scriptsize\mbox{fi}}=\left(C^{\scriptsize\mbox{fi}}_1, \ldots,C^{\scriptsize\mbox{fi}}_{n^{\scriptsize\mbox{fi}}-1}\right),
 $$
 where $C^{\scriptsize\mbox{fi}}_j=\left(s(x^{\scriptsize\mbox{fi}}_j),s(x^{\scriptsize\mbox{fi}}_j+1), \ldots,s(x^{\scriptsize\mbox{fi}}_{j+1})\right)$ is
 the $j$th cycle of the finer cuts $(x^{\scriptsize\mbox{fi}}_i,y^{\scriptsize\mbox{fi}}_i)$ with $i=1,\ldots,n^{\scriptsize\mbox{fi}}$  and $j=1,\ldots,n^{\scriptsize\mbox{fi}}-1$. Moreover, the $j$th cycle cuts are $(x^{\scriptsize\mbox{fi}}_j,y^{\scriptsize\mbox{fi}}_j)$ and $(x^{\scriptsize\mbox{fi}}_{j+1},y^{\scriptsize\mbox{fi}}_{j+1})$. And the lengths of cycles are
 $$
 D^{\scriptsize\mbox{fi}}=\left(d^{\scriptsize\mbox{fi}}_1, \ldots,d^{\scriptsize\mbox{fi}}_{n^{\scriptsize\mbox{fi}}-1}\right),
 $$
 where $d^{\scriptsize\mbox{fi}}_j=x^{\scriptsize\mbox{fi}}_{j+1}-x^{\scriptsize\mbox{fi}}_{j}$ is the $j$the cycle's length with $j=1,\ldots,n^{\scriptsize\mbox{fi}}-1$. The best cycle $C^{\scriptsize\mbox{be}}$ is the cycle best represents the signal's gait features in the sense of statistical distribution; it will be used as a reference to find the rest of cycles on the signal. Let $H^{\scriptsize\mbox{fi}}=\left(h^{\scriptsize\mbox{fi}}_1, \ldots,h^{\scriptsize\mbox{fi}}_{n^{\scriptsize\mbox{fi}}-1}\right)$ and $E^{\scriptsize\mbox{fi}}=\left(e^{\scriptsize\mbox{fi}}_1, \ldots,e^{\scriptsize\mbox{fi}}_{n^{\scriptsize\mbox{fi}}}\right)$ be the values of histogram and the bin edges of $D^{\scriptsize\mbox{fi}}$ respectively. Let $d^{\scriptsize\mbox{be}}$ be the length of the best cycle $C^{\scriptsize\mbox{be}}$.
 
 We are going to find the best cycle on the signal in at most three steps. See Algorithm \ref{best_cycle_step1} for details.
 \begin{enumerate}
     \item[Step 1:] Attempt to search for the best cycle that satisfies
     \begin{itemize}
         \item A cycle has length that is among the non-zero frequency cycle lengths of the finer cuts.
         \item The length of the cycle needs to be close to the hypothesized gait length $d^{\scriptsize\mbox{hy}}$, where hypothesized gait is a hyper-parameter to be tuned or given. Normally $d^{\scriptsize\mbox{hy}}$ is the average gait length, but it can be modified to obtain different goals such as half cycle segmentation.
     \end{itemize}
     If the two requirements of a best cycle can be satisfied by any cycle of the finer cuts, we will use that cycle as the best cycle and skip the next step. Otherwise, go to the next step.
     \item[Step 2:] In this step, we are going to modify the finer cuts based on two situations below to achieve the requirements in Step 1.
     \begin{itemize}
         \item If $d^{\scriptsize\mbox{be}}$ is much smaller than the gait length $d^{\scriptsize\mbox{hy}}$, we are going to combine each nearby pair of the finer cuts to make the average cycle length of finer cuts larger than before.
         \item If $d^{\scriptsize\mbox{be}}$ is much larger than the gait length $d^{\scriptsize\mbox{hy}}$, we are going back to Session \ref{finer_processing} and pick up more finer cuts.
     \end{itemize}
     Either of the two modifications will renew the finer cuts. Then search for the best cycle on the new finer cuts based on the following conditions:
     \begin{itemize}
         \item A cycle has length that is closest to the largest frequency of cycle lengths of the finer cuts.
         \item The length of the cycle needs to be close to the hypothesized gait length $d^{\scriptsize\mbox{hy}}$.
     \end{itemize}
     If no best cycle obtained, repeat Step 2 until one is found.
 \end{enumerate}
Up to this point, we have found the best cycle $C^{\scriptsize\mbox{be}}$ on the signal. Next, we will discuss how to find the rest of cycles on the signal based on the features of $C^{\scriptsize\mbox{be}}$.

\begin{algorithm}
\SetAlgoLined
\KwOut{$C^{\scriptsize\mbox{be}}$}
\KwIn{$(x^{\scriptsize\mbox{fi}}_i,y^{\scriptsize\mbox{fi}}_i), i=1,\ldots,n^{\scriptsize\mbox{fi}},  d^{\scriptsize\mbox{hy}} $}
$(x^{\scriptsize\mbox{fi}}_i,y^{\scriptsize\mbox{fi}}_i), i=1,\ldots,n_{\scriptsize\mbox{fi}}\implies
D^{\scriptsize\mbox{fi}},H^{\scriptsize\mbox{fi}}, E^{\scriptsize\mbox{fi}}$\;
 \For{$i=1,\ldots, n^{\scriptsize\mbox{fi}}-1$ }{
  $d^{\scriptsize\mbox{be}}=\left(e^{\scriptsize\mbox{fi}}_{i+1}-e^{\scriptsize\mbox{fi}}_{i}\right)/2$\;
  \If{$h^{\scriptsize\mbox{fi}}_{i}==0$}{
   $d^{\scriptsize\mbox{be}}=0$\;
   break\;
   }
  \If{$0.8\times d^{\scriptsize\mbox{hy}}\leq d^{\scriptsize\mbox{be}}\leq1.2\times d^{\scriptsize\mbox{hy}}$}{
   break\;
   }}
   n=1\;
  \While{$0.8\times d^{\scriptsize\mbox{be}}\leq d^{\scriptsize\mbox{hy}}$ or $d^{\scriptsize\mbox{be}}\geq1.2\times d^{\scriptsize\mbox{hy}}$}{
  
  \eIf{$d^{\scriptsize\mbox{be}}\leq0.8\times d^{\scriptsize\mbox{hy}}$}{
  $(x^{\scriptsize\mbox{fi}}_i,y^{\scriptsize\mbox{fi}}_i)\leftarrow(x^{\scriptsize\mbox{fi}}_i,y^{\scriptsize\mbox{fi}}_i)$ for $i= \left\{\begin{array}{ll}
    1,3,5,\ldots,n^{\scriptsize\mbox{fi}}-1   & \mbox{if}~ n^{\scriptsize\mbox{fi}}-1 ~\mbox{is odd} \\
    1,3,5,\ldots,n^{\scriptsize\mbox{fi}}-2   & \mbox{if}~ n^{\scriptsize\mbox{fi}}-1 ~\mbox{is even}
  \end{array}\right.$\;
  $(x^{\scriptsize\mbox{fi}}_i,y^{\scriptsize\mbox{fi}}_i), i=1,3,5,\ldots,n^{\scriptsize\mbox{fi}}-1~\mbox{or}~ n^{\scriptsize\mbox{fi}}-2\implies
  D^{\scriptsize\mbox{fi}},H^{\scriptsize\mbox{fi}}, E^{\scriptsize\mbox{fi}}$\;
  $i_{\scriptsize\max} = \operatorname{argmax}(h^{\scriptsize\mbox{fi}}_i)$\;
  $d^{\scriptsize\mbox{be}}=\left(e^{\scriptsize\mbox{fi}}_{i_{\scriptsize\max}+1}-e^{\scriptsize\mbox{fi}}_{i_{\scriptsize\max}+1}\right)/2$\;
  }{
  \If{n==6}{$d^{\scriptsize\mbox{be}}=d^{\scriptsize\mbox{hy}}$\;break\;}
  $(x^{\scriptsize\mbox{fi}}_i,y^{\scriptsize\mbox{fi}}_i)\leftarrow$ Repeat Section \ref{finer_processing} but pick peaks with angles above $(0.5-n\times0.1)$th quantile\;
  $(x^{\scriptsize\mbox{fi}}_i,y^{\scriptsize\mbox{fi}}_i)\implies D^{\scriptsize\mbox{fi}}, H^{\scriptsize\mbox{fi}}, E^{\scriptsize\mbox{fi}}$ and $i$ has new indexes\;
  $i_{\scriptsize\max} = \operatorname{argmax}(h^{\scriptsize\mbox{fi}}_i)$\;
  $d^{\scriptsize\mbox{be}}=\left(e^{\scriptsize\mbox{fi}}_{i_{\scriptsize\max}+1}-e^{\scriptsize\mbox{fi}}_{i_{\scriptsize\max}+1}\right)/2$\;
  $n=n+1$
  }
  $i_{\scriptsize\min} =\operatorname{argmin}(\left|d^{\scriptsize\mbox{fi}}_i-d^{\scriptsize\mbox{be}}\right|)$\;
  $C^{\scriptsize\mbox{be}}=\left(s(x^{\scriptsize\mbox{fi}}_{i_{\scriptsize\min}}),s(x^{\scriptsize\mbox{fi}}_{i_{\scriptsize\min}}+1), \ldots,s(x^{\scriptsize\mbox{fi}}_{i_{\scriptsize\min}+1})\right)$.
  }
\caption{Find best cycle}
\label{best_cycle_step1}
\end{algorithm}
 
\subsection{Find all cycles} 
\label{find_all_cycles}
  It remains to cut the rest of the signal into cycles to obtain full cycle segmentation, which is denoted by 
  $$
  \mathbf C^{\scriptsize\mbox{se}}=\left(C^{\scriptsize\mbox{se}}_1,\ldots,C^{\scriptsize\mbox{be}},\ldots,C^{\scriptsize\mbox{se}}_M\right),
  $$
  where $C^{\scriptsize\mbox{se}}_m=\left(s(x^{\scriptsize\mbox{se}}_{m}), \ldots,s(x^{\scriptsize\mbox{se}}_{m+1})\right)$  and the cutting positions are 
  $$
  X^{\scriptsize\mbox{se}}=\left(x^{\scriptsize\mbox{se}}_1,\ldots,x^{\scriptsize\mbox{se}}_{M+1}\right)
  $$
  for $m=1,\ldots,M$. In this section, we mainly introduce the approach of finding other cycles to the left of the best cycle $C^{\scriptsize\mbox{be}}=\left(s(x^{\scriptsize\mbox{fi}}_{i_{\scriptsize\min}}),s(x^{\scriptsize\mbox{fi}}_{i_{\scriptsize\min}}+1), \ldots,s(x^{\scriptsize\mbox{fi}}_{i_{\scriptsize\min}+1})\right)$, and the right-hand side cycles can be found easily by a similar approach, so it is omitted. Two steps are included in searching to the left:
  \begin{enumerate}
      \item Find cycles based on the left-hand side finer-picked peaks. Since the finer cuts are highly possible the right cutting positions, it deserves to scan one by one of these cuts to utilize the previously calculated results. For instance, the left-hand side peak right next to $x^{\scriptsize\mbox{fi}}_{i_{\scriptsize\min}}$ is $x^{\scriptsize\mbox{fi}}_{i_{\scriptsize\min}-1}$. Let the potential cut position on the finer cuts be $p^{\scriptsize\mbox{po}}$ and  $p^{\scriptsize\mbox{po}}$ is now at the $i_{\scriptsize\min}$th cut of finer cuts, or $p^{\scriptsize\mbox{po}}=i_{\scriptsize\min}$. So the potential cycle length is $d^{\scriptsize\mbox{po}}=x^{\scriptsize\mbox{fi}}_{p^{\scriptsize\mbox{po}}}-x^{\scriptsize\mbox{fi}}_{p^{\scriptsize\mbox{po}}-1}=x^{\scriptsize\mbox{fi}}_{i_{\scriptsize\min}}-x^{\scriptsize\mbox{fi}}_{i_{\scriptsize\min}-1}$ and the left over signal length is $d^{\scriptsize\mbox{le}}=0$. There are three conditions listed in the following:
      \begin{itemize}
          \item If $d^{\scriptsize\mbox{po}}+d^{\scriptsize\mbox{le}}$ is close to $d^{\scriptsize\mbox{be}}$, the two finer cuts are kept as a cycle, i,e, $X^{\scriptsize\mbox{se}}=\left(x^{\scriptsize\mbox{fi}}_{i_{\scriptsize\min}-1},x^{\scriptsize\mbox{fi}}_{i_{\scriptsize\min}}, x^{\scriptsize\mbox{fi}}_{i_{\scriptsize\min}+1}\right)$. Then update $p^{\scriptsize\mbox{po}}=i_{\scriptsize\min}-1$ and $d^{\scriptsize\mbox{le}}=0$ for next iteration.
          \item If $d^{\scriptsize\mbox{po}}+d^{\scriptsize\mbox{le}}$ is much smaller  than $d^{\scriptsize\mbox{be}}$, $x^{\scriptsize\mbox{fi}}_{i_{\scriptsize\min}-1}$ is disregarded and update $d^{\scriptsize\mbox{le}}=x^{\scriptsize\mbox{fi}}_{i_{\scriptsize\min}}-x^{\scriptsize\mbox{fi}}_{i_{\scriptsize\min}-1}$ and $p^{\scriptsize\mbox{po}}=i_{\scriptsize\min}-1$ for next iteration.
          \item If $d^{\scriptsize\mbox{po}}+d^{\scriptsize\mbox{le}}$ is much larger than $d^{\scriptsize\mbox{be}}$, then find the possible cycle in between the two finer cuts and update to $X^{\scriptsize\mbox{se}}$ . Then update $p^{\scriptsize\mbox{po}}=i_{\scriptsize\min}-1$ and record the left over signal length $d^{\scriptsize\mbox{le}}=x^{\scriptsize\mbox{se}}_{1}-x^{\scriptsize\mbox{fi}}_{i_{\scriptsize\min}-1}$.
      \end{itemize}
    \item Since the finer cuts may not reach the very left end of the signal, we are going to keep searching for cycles left to the finer cuts if there is any. 
  \end{enumerate}
 The above steps are iteratively repeated until the end of the signal is reached. There are two hyper-parameters $\beta^{\scriptsize\mbox{hy}}$ and $l^{\scriptsize\mbox{hy}}$  involved to be given or tuned where $\beta^{\scriptsize\mbox{hy}}$ controls how similar the two cycles are and  $l^{\scriptsize\mbox{hy}}$ controls the range of searching for local minimum. Details will be shown in Algorithm \ref{find_cycle_on_the_left}. 
  
\begin{algorithm}
\SetAlgoLined
\KwOut{$\mathbf C^{\scriptsize\mbox{se}}$}
\KwIn{$(x^{\scriptsize\mbox{fi}}_i,y^{\scriptsize\mbox{fi}}_i), i=1,\ldots,n^{\scriptsize\mbox{fi}},  d^{\scriptsize\mbox{be}}, C^{\scriptsize\mbox{be}}  $}
$p^{\scriptsize\mbox{po}}=i_{\scriptsize\min}$\;
$d^{\scriptsize\mbox{le}}=0$\;
$\mathbf{C}^{\scriptsize\mbox{se}}=\left(s(x^{\scriptsize\mbox{fi}}_{p^{\scriptsize\mbox{po}}}), \ldots,s(x^{\scriptsize\mbox{fi}}_{p^{\scriptsize\mbox{po}}+1})\right)$\;
 \While{$p^{\scriptsize\mbox{po}}>0$ }{
  $d^{\scriptsize\mbox{be}}=\frac{1}{M}\sum_{m=1}^M\left(x^{\scriptsize\mbox{se}}_{m+1}-x^{\scriptsize\mbox{se}}_m\right)$ with $\mathbf{C}^{\scriptsize\mbox{se}}=\left(C^{\scriptsize\mbox{se}}_1,\ldots,C^{\scriptsize\mbox{se}}_M\right)$ and $C^{\scriptsize\mbox{se}}_m=\left(s(x^{\scriptsize\mbox{se}}_{m}), \ldots,s(x^{\scriptsize\mbox{se}}_{m+1})\right)$\;
  $d^{\scriptsize\mbox{po}}=x^{\scriptsize\mbox{fi}}_{p^{\scriptsize\mbox{po}}}-x^{\scriptsize\mbox{fi}}_{p^{\scriptsize\mbox{po}}-1}$\;
  $C^{\scriptsize\mbox{po}}=\left(s(x^{\scriptsize\mbox{fi}}_{p^{\scriptsize\mbox{po}}-1}),\ldots,s(x^{\scriptsize\mbox{fi}}_{p^{\scriptsize\mbox{po}}})\right)$\;
  \uIf{$0.95\times d^{\scriptsize\mbox{be}}\leq d^{\scriptsize\mbox{po}}+d^{\scriptsize\mbox{le}}\leq1.05\times d^{\scriptsize\mbox{be}} $}{
   \If{The correlation of $C^{\scriptsize\mbox{po}}$ and $C^{\scriptsize\mbox{se}}_1$ is less than $\beta^{\scriptsize\mbox{hy}}$ }{break\;}
   $\mathbf{C}^{\scriptsize\mbox{se}}\leftarrow\left(C^{\scriptsize\mbox{po}}, C^{\scriptsize\mbox{se}}\right)$\;
   $d^{\scriptsize\mbox{le}}=0$\;
   }
   \uElseIf{$ d^{\scriptsize\mbox{po}}+d^{\scriptsize\mbox{le}}\leq0.95\times d^{\scriptsize\mbox{be}} $}{
    $d^{\scriptsize\mbox{le}}\leftarrow d^{\scriptsize\mbox{po}}+d^{\scriptsize\mbox{le}}$\;
    }
    \Else{$x^{\scriptsize\mbox{in}}_1, \ldots,x^{\scriptsize\mbox{in}}_{n^{\scriptsize\mbox{po}}}$ or $x^{\scriptsize\mbox{fi}}_{p^{\scriptsize\mbox{po}}-1}=$ positions that divide $x^{\scriptsize\mbox{fi}}_{p^{\scriptsize\mbox{po}}}-x^{\scriptsize\mbox{fi}}_{p^{\scriptsize\mbox{po}}-1}$ from the right into pieces with length $d^{\scriptsize\mbox{be}}$\;
    \For{$i=n^{\scriptsize\mbox{po}}-1,\ldots,1$}
    {$x^{\scriptsize\mbox{in}}_i\leftarrow\operatorname{argmin}_{j\in[x^{\scriptsize\mbox{in}}_i-l^{\scriptsize\mbox{hy}}, x^{\scriptsize\mbox{in}}_i+l^{\scriptsize\mbox{hy}}]}s(j)$\;
    \If{The correlation of $\left(s(x^{\scriptsize\mbox{in}}_i), \ldots, s(x^{\scriptsize\mbox{in}}_{i+1})\right)$ and $C^{\scriptsize\mbox{se}}_1$ is less than $\beta^{\scriptsize\mbox{hy}}$ }{break\;}
    $\mathbf{C}^{\scriptsize\mbox{se}}\leftarrow\left(\left(s(x^{\scriptsize\mbox{in}}_i), \ldots, s(x^{\scriptsize\mbox{in}}_{i+1})\right), \mathbf{C}^{\scriptsize\mbox{se}}\right)$ \;
    }
    
    $d^{\scriptsize\mbox{le}}=x^{\scriptsize\mbox{se}}_{1}-x^{\scriptsize\mbox{fi}}_{p^{\scriptsize po}-1}$
    }
    $p^{\scriptsize po}\leftarrow p^{\scriptsize po}-1$
  }
  \While{$s(x^{\scriptsize\mbox{se}}_{1})-s(1)\geq d^{\scriptsize\mbox{be}}$}{
  $d^{\scriptsize\mbox{be}}=\frac{1}{M}\sum_{m=1}^M\left(x^{\scriptsize\mbox{se}}_{m+1}-x^{\scriptsize\mbox{se}}_m\right)$ with $\mathbf{C}^{\scriptsize\mbox{se}}=\left(C^{\scriptsize\mbox{se}}_1,\ldots,C^{\scriptsize\mbox{se}}_M\right)$ and $C^{\scriptsize\mbox{se}}_m=\left(s(x^{\scriptsize\mbox{se}}_{m}), \ldots,s(x^{\scriptsize\mbox{se}}_{m+1})\right)$\;
  $x^{\scriptsize\mbox{in}}_0= x^{\scriptsize\mbox{se}}_{1}-d^{\scriptsize\mbox{be}}$\;
  $x^{\scriptsize\mbox{in}}_0\leftarrow\operatorname{argmin}_{j\in[x^{\scriptsize\mbox{in}}-l^{\scriptsize\mbox{hy}}, x^{\scriptsize\mbox{in}}+l^{\scriptsize\mbox{hy}}]}s(j)$\;
  \If{The correlation of $\left(s(x^{\scriptsize\mbox{in}}_0), \ldots, s(x^{\scriptsize\mbox{se}}_1)\right)$ and $C^{\scriptsize\mbox{se}}_1$ is less than $\beta^{\scriptsize\mbox{hy}}$ }{break\;}
  $\mathbf{C}^{\scriptsize\mbox{se}}\leftarrow\left(\left(s(x^{\scriptsize\mbox{in}}_0), \ldots, s(x^{\scriptsize\mbox{se}}_{1})\right), \mathbf{C}^{\scriptsize\mbox{se}}\right)$ \;
  }
\caption{Find cycles on the left of $C^{\scriptsize\mbox{be}}$}
\label{find_cycle_on_the_left}
\end{algorithm}  

\section{ New signal distance}
\label{signal_distance}
We propose a new way of calculating distance between two signals. Let $S_1 = (s_1(x_1),\ldots, s_1(x_{n_{1}}))'$ and $S_2 = (s_2(y_1),\ldots, s_2(y_{n_{2}}))'$ be two signals and $\left(x_i, s_1(x_i)\right), i=1,\ldots,n_1$ and $\left(y_j, s_2(y_j)\right), j=1,\ldots,n_2$ are points on $S_1$ and $S_2$ respectively. Let $z=\left(z_1, \ldots, z_{n_1+n_2}\right)$ be a permutation of $\left(x_1, \ldots, x_{n_1}, y_1,\ldots, y_{n_2}\right)$ such that 
$$
z_1\leq z_2\leq\ldots\leq z_{n_1+n_2}.
$$
The basic idea is to find the interpolations of $z$ on $S_1$ and $S_2$, denoted by $s_1(z_h)$ and $s_2(z_h)$ with $h=1,\ldots, n_1+n_2$.
Then the distance between $S_1$ and $S_2$ is defined as 
$$
\mathcal{D}(S_1, S_2)=\left(\sum_{h=1}^{n_1+n_2}(s_1(z_h)-s_2(z_h))^2\right)^{1/2}.
$$
Note that here we use Euclidean distance to formalize $\mathcal{D}(S_1, S_2)$; in fact any distance can be fitted here depends on different situations. The details are shown in Algorithm \ref{signal_distance}. This signal distance plays a crucial role in all signal analysis. In particularly, we are going to show how it will be applied in segmentation and human motion identification.

\begin{algorithm}[H]
\SetAlgoLined
\KwOut{$\mathcal{D}(S_1, S_2)$}
\KwIn{$S_1, S_2$}
$S_1, S_2\implies
 z$\;
 \For{$h=1,\ldots, n_1+n_2$ }{
  $s_1(z_h)=\left\{\begin{array}{ll}
    s_1(x_i)+\left(s_1(x_{i+1})-s_1(x_{i})\right)\frac{z_h-x_i}{x_{i+1}-x_i}  &\mbox{if~}\exists  i\in\{1,\ldots,n_1\} \mbox{~such that~} x_i\leq z_h\leq x_{i+1}  \\
    s_1(x_1)   & \mbox{if~} z_h\leq x_{1}\\
    s_1(x_{n_1})& \mbox{if~} z_h\geq x_{n_1}
  \end{array}\right.$\;
  $s_2(z_h)=\left\{\begin{array}{ll}
    s_2(y_j)+\left(s_2(y_{j+1})-s_2(y_{j})\right)\frac{z_h-y_j}{y_{j+1}-y_j}  &\mbox{if~}\exists  j\in\{1,\ldots,n_2\} \mbox{~such that~} y_j\leq z_h\leq y_{j+1}  \\
    s_2(y_1)   & \mbox{if~} z_h\leq y_{1}\\
    s_2(y_{n_2})& \mbox{if~} z_h\geq y_{n_2}
  \end{array}\right.$;
  }
  $\mathcal{D}(S_1, S_2)=\left(\sum_{h=1}^{n_1+n_2}(s_1(z_h)-s_2(z_h))^2\right)^{1/2}.$
\caption{Signal distance}
\label{signal_distance}
\end{algorithm}

\section{Experimental study}\label{experimental}
All the experimental results in this section are based on HAPT data \footnote{http://archive.ics.uci.edu/ml/datasets/Smartphone-Based+Recognition+of+Human+Activities+and+Postural+Transitions\#}. There are 30 volunteers in total within an age bracket of 19-48 years engaged in the data recording and all were wearing a smartphone (Samsung Galaxy S II) on the waist and performed a protocol of activities during the experiment execution. Among all the activities, walking, walking downstairs and walking upstairs are mainly studied in this paper. More specifically, 3-axial accelerometer and gyroscope data for each of the activity are studied. 

The segmentation in Section \ref{segmentation} shows the results of cutting on the data set by using algorithm proposed in Section \ref{segmentation_algorithm}. The user-adversary identification in Section \ref{user_adversary_identification} will continue use the segmented data. Training data are clustered to extract individual archetypes or features, while each testing cycle is compared with all volunteer's archetypes extracted in training process to do classification. Each volunteer may have several separated periods of data recorded for each activity; several periods are studied to capture full picture of the archetypes for an individual.

\subsection{Segmentation}
\label{segmentation}
The segmentation algorithm discussed in Section \ref{segmentation_algorithm} is applied in this section on the HAPT data set. Recall that there are three hyper-parameters involved in the segmentation algorithm: hypothesized gait length $d^{\scriptsize\mbox{hy}}$,   cycle similarity threshold $\beta^{\scriptsize\mbox{hy}}$ and local minimum searching range $l^{\scriptsize\mbox{hy}}$. The segmentation structure tunes the positions of the cuts on the signal by the three hyper-parameters on a score defined as
\begin{equation*}
   \mathcal C\left(d^{\scriptsize\mbox{hy}}, \beta^{\scriptsize\mbox{hy}}, l^{\scriptsize\mbox{hy}} \right)=\sum_{i=1}^{M-1}\min_{n\in[n_1,n_2]}\mathcal{D}(\left(s(x^{\scriptsize\mbox{se}}_i+n), \ldots, s(x^{\scriptsize\mbox{se}}_{i+1}+n)\right), \left(s(x^{\scriptsize\mbox{se}}_{i+1}), \ldots, s(x^{\scriptsize\mbox{se}}_{i+2})\right)),
\end{equation*}
where $n$ is a shifting parameter that can bring the two signal $\left(s(x^{\scriptsize\mbox{se}}_i), \ldots, s(x^{\scriptsize\mbox{se}}_{i+1})\right)$ and $\left(s(x^{\scriptsize\mbox{se}}_{i+1}), \ldots, s(x^{\scriptsize\mbox{se}}_{i+2})\right))$ as close as possible, and in our case, $n$ should be in the range of $[\left(x^{\scriptsize\mbox{se}}_{i+2}-x^{\scriptsize\mbox{se}}_{i}\right)/2-5, \left(x^{\scriptsize\mbox{se}}_{i+2}-x^{\scriptsize\mbox{se}}_{i}\right)/2+5]$. Eventually, the segmentation we obtained is $ \mathbf C^{\scriptsize\mbox{se}}=\operatorname{argmin}_{d^{\scriptsize\mbox{hy}}, \beta^{\scriptsize\mbox{hy}}, l^{\scriptsize\mbox{hy}}}\mathcal C\left(d^{\scriptsize\mbox{hy}}, \beta^{\scriptsize\mbox{hy}}, l^{\scriptsize\mbox{hy}} \right)$. Based on our basic statistical analysis, here for a full gait cycle, $d^{\scriptsize\mbox{hy}}$ is normally around 50, which follows the value of $l^{\scriptsize\mbox{hy}}$ less than $20$. We are going to set $ \beta^{\scriptsize\mbox{hy}}<0$ to cut all possible cycles in the signal.

\begin{figure}[h]
\centering
\begin{subfigure}{\columnwidth}
\centering
\includegraphics[width=0.8\textwidth]{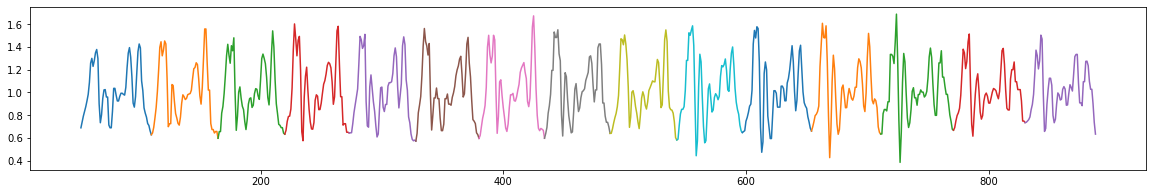}
\caption{}
\end{subfigure}
\begin{subfigure}{\columnwidth}
\centering
\includegraphics[width=0.8\textwidth]{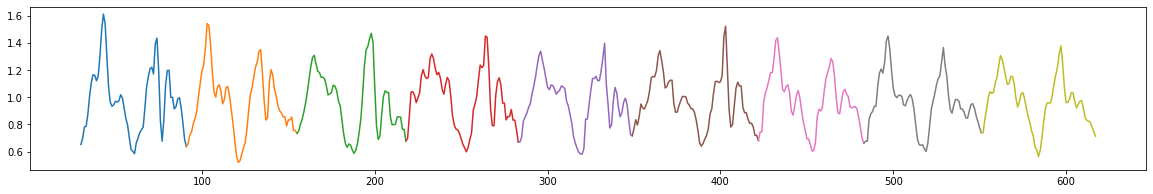}
\caption{}
\end{subfigure}
\begin{subfigure}{\columnwidth}
\centering
\includegraphics[width=0.8\textwidth]{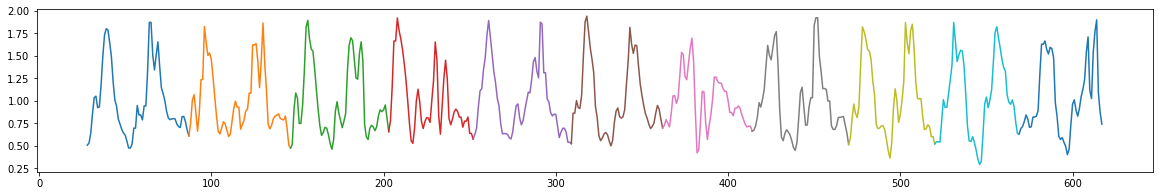}
\caption{}
\end{subfigure}
\caption{X-axis full accelerometer segmentation on 1st period of volunteer 1 in HAPT data set with different activities. Top (a): walking; middle (b): walking up; bottom (c): walking down.}
\label{full_seg_1}
\end{figure}

\begin{figure}[h]
\centering
\begin{subfigure}{\columnwidth}
\centering
\includegraphics[width=0.8\textwidth]{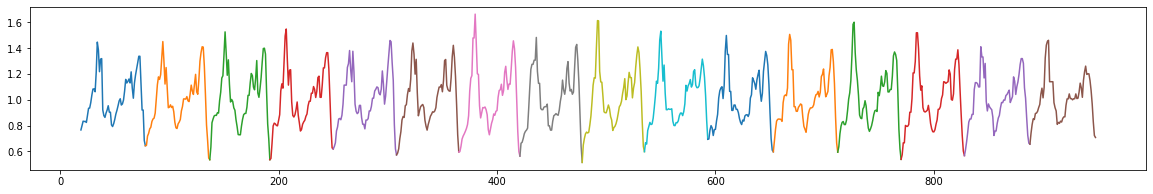}
\caption{}
\end{subfigure}
\begin{subfigure}{\columnwidth}
\centering
\includegraphics[width=0.8\textwidth]{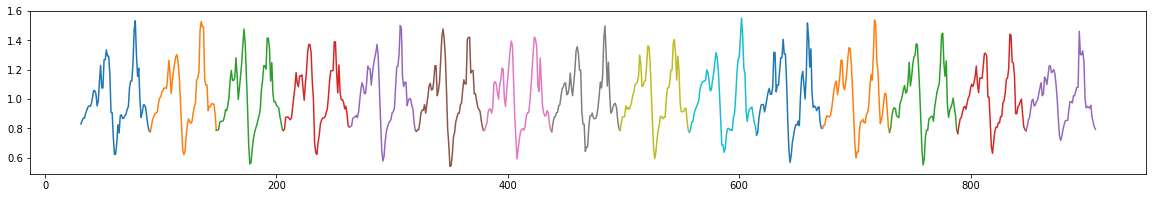}
\caption{}
\end{subfigure}
\begin{subfigure}{\columnwidth}
\centering
\includegraphics[width=0.8\textwidth]{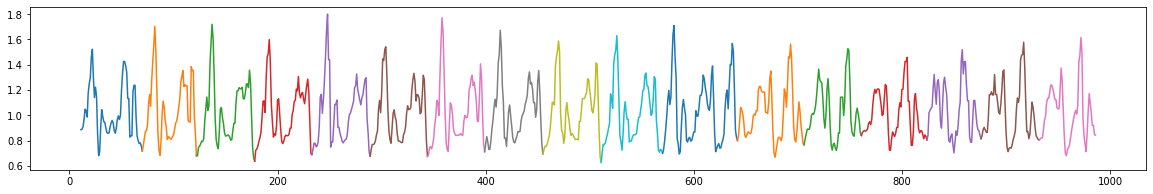}
\caption{}
\end{subfigure}
\caption{X-axis full accelerometer segmentation for walking. Top (a): period 1 of volunteer 15 in HAPT data set; middle (b): period 2 of volunteer 15 in HAPT data set; bottom (c): period 2 of volunteer 27 in HAPT data set. }
\label{full_seg_2}
\end{figure}

We apply the segmentation on  the x-axis accelerometer of walking, walking up, walking down and mixed walking activities for each volunteer. Some results are included in Figure \ref{full_seg_1} and Figure \ref{full_seg_2}. Figure \ref{full_seg_1} indicates that the segmentation algorithm works for different activities: walking, walking up and walking down. All the periods of walking, walking up and walking down in the HAPT data set can be perfectly cut by our proposed approach, except for the case shown in Figure \ref{full_seg_2}. 

\begin{figure}[h]
\centering
\begin{subfigure}{\columnwidth}
\centering
\includegraphics[width=0.8\textwidth]{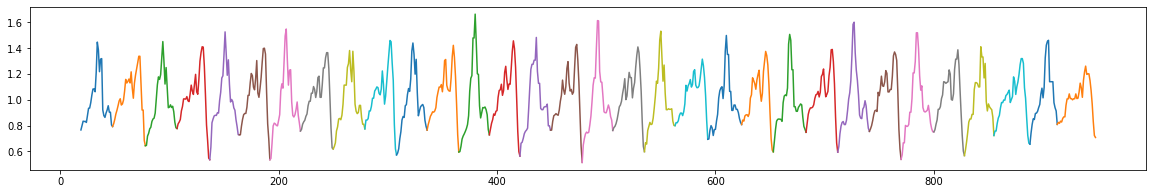}
\caption{}
\end{subfigure}
\begin{subfigure}{\columnwidth}
\centering
\includegraphics[width=0.8\textwidth]{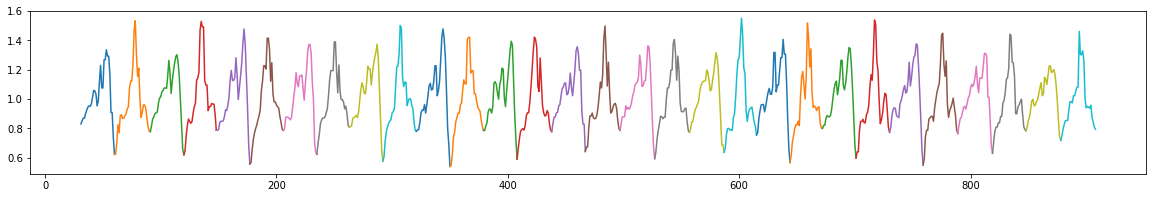}
\caption{}
\end{subfigure}
\begin{subfigure}{\columnwidth}
\centering
\includegraphics[width=0.8\textwidth]{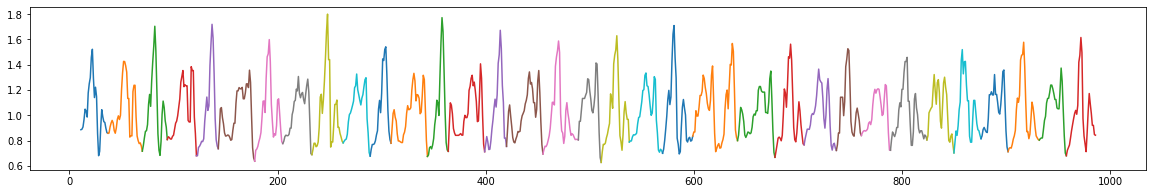}
\caption{}
\end{subfigure}
\begin{subfigure}{\columnwidth}
\centering
\includegraphics[width=0.8\textwidth]{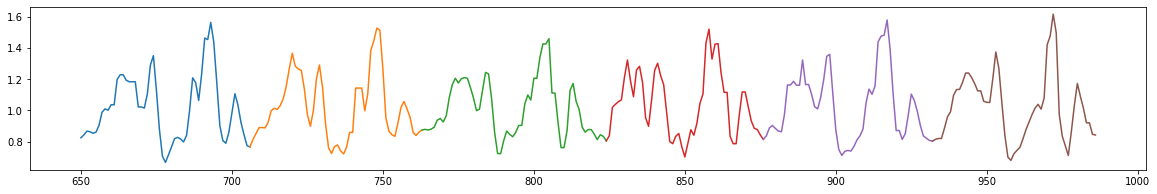}
\caption{}
\end{subfigure}
\caption{X-axis accelerometer  segmentation for walking. 1st plot (a): half segmentation of period 1 of volunteer 15 in HAPT data set; 2nd plot (b): half segmentation of period 2 of volunteer 15 in HAPT data set; 3rd plot (c): half segmentation of period 2 of volunteer 27 in HAPT data set; 4th plot (d): full segmentation of period 2 of volunteer 27 in HAPT data set with positive $\beta^{\scriptsize\mbox{hy}}$. }
\label{half_seg_1}
\end{figure}

Like what is shown in top and middle figures in Figure \ref{full_seg_2}, depending on which leg step forward first in the experiments, the segmentation might not be consistent. Even though we can say both of them are perfectly segmentation individually, it might become a problem in applications like user-adversary identification in next section. Our way of addressing this problem in the user-adversary identification process is to include each of the different cutting as a archetype for the individual. Besides, this problem can also be easily resolved by half segmentation of the gait cycle. More precisely, we can easily set $d^{\scriptsize\mbox{hy}}$ smaller that that of full cycle and go through the same process. See Figure \ref{half_seg_1} first and second plots. 

Another problem is shown in bottom figure in Figure \ref{full_seg_2}. Although certain period of certain activity is recorded at the same time with same volunteer, it sometimes not that stable. In this example, the first $2/3$ signal looks very different from the last $1/3$ of it, resulting in some errors in segmentation. Again, half segmentation can also resolve this problem, see Figure  \ref{half_seg_1} third plot. Another alternative is using bigger $\beta^{\scriptsize\mbox{hy}}$ which guarantee the similarity between cycles are maintained , see Figure  \ref{half_seg_1} last plot. This alternative is more feasible for applications that detect the walking behaviors by cutting off the non-walking signals. The way we deal with this problem in next section is to repeat the latter alternative to extract different patterns of the gait cycle as more archetypes for the individual user.

\begin{figure}[h]
\centering
\begin{subfigure}{\columnwidth}
\centering
\includegraphics[width=0.8\textwidth]{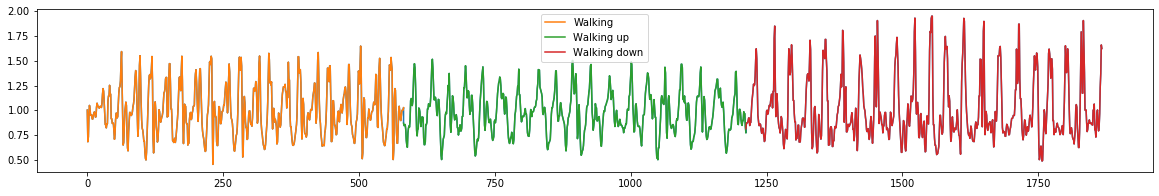}
\caption{}
\end{subfigure}
\begin{subfigure}{\columnwidth}
\centering
\includegraphics[width=0.8\textwidth]{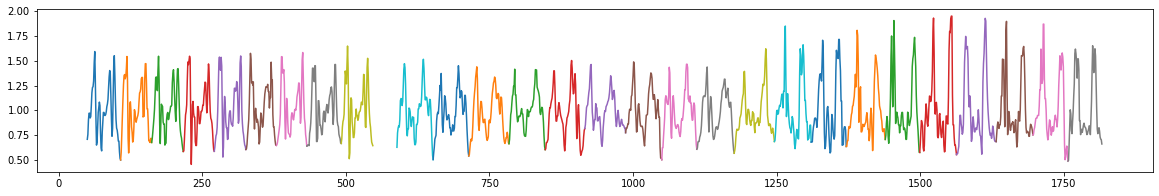}
\caption{}
\end{subfigure}
\caption{X-axis accelerometer segmentation for mixed walking signal of volunteer 1 in HAPT data set. Top (a): manually combined mixed walking signal; bottom (b): iterative segmentation results. }
\label{mix_seg}
\end{figure}

Our segmentation algorithm also works for a more complicated situation  when all types of walking cycles are mixed. By iteratively running our segmentation algorithm, we can separately segment the different types of walking signals and cut off the non-walking or noise signal. For example, we manually combined X-axis accelerometer walking, walking up and walking down signal of volunteer 1 in HAPT data set and then we apply our proposed algorithm to the new mixed data set; the segmentation results is shown in Figure \ref{mix_seg}.

\subsection{User-adversary identification}
\label{user_adversary_identification}
\begin{figure}
    \centering
    \includegraphics[width=0.5\textwidth]{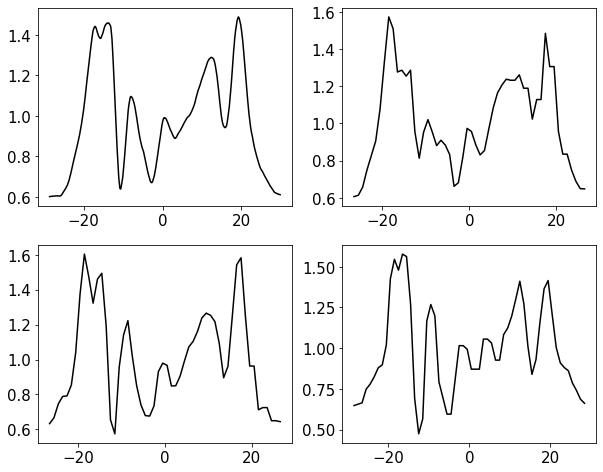}
    \caption{Archetypes volunteer 1 based on x-axis acceleromater signal; the threshold $\varrho=0.27$.}
    \label{arthetype_example}
\end{figure}

In this section, multi-class classification is processed to identify user and adversary based on the well-cut x-axis accelerometer based gait segmentation from Section \ref{segmentation}. The cut cycles are split into training (80\%) and testing (20\%) set  for each volunteer. Cycles in the training set are used to  extract the archetypes for each individual. The approach to obtain the archetypes is shown in Algorithm \ref{native_clustering} and that the main idea is about clustering the training cycles into classes such that cycles within each class are guaranteed to have signal distance less than a threshold $\varrho$. The threshold $\varrho$ controls the speed and accuracy of the classification process; a high $\varrho$ speeds up the process but results in lower accuracy. Archetypes are obtained by taking "average" of the cycles in each classes using the new signal distance defined. As a result, each volunteer will have several archetypes, denoted as
  $\mathbf C^{\scriptsize\mbox{cl}}=\left(C^{\scriptsize\mbox{cl}}_1,\ldots,C^{\scriptsize\mbox{cl}}_{M^{\scriptsize\mbox{cl}}}\right),
  $
  where $C^{\scriptsize\mbox{cl}}_m=\left(s(x^{\scriptsize\mbox{cl}}_{m}), \ldots,s(x^{\scriptsize\mbox{cl}}_{m+1})\right)$. See Figure \ref{arthetype_example} for an example of archetypes of volunteer 1 with $\varrho=0.27$. Note that all the experiments in this paper use $\varrho=0.1$ to obtain more archetypes and then higher performance. 
It is important to mention that, each test sample is single cycle based in the testing process; it is labeled into the a volunteer that has the archetype closed to the test cycle by using the signal distance. Our single cycle test method is in demand in cybersecurity field since user authentication has to be done within a few seconds.


\begin{algorithm}
\SetAlgoLined
\KwOut{$\mathbf{C}^{\scriptsize\mbox{cl}}=\left(C^{\scriptsize\mbox{cl}}_1,\ldots,C^{\scriptsize\mbox{cl}}_{M^{\scriptsize\mbox{cl}}}\right)$}
\KwIn{$\mathbf{C}^{\scriptsize\mbox{se}}=\left(C^{\scriptsize\mbox{se}}_1,\ldots,C^{\scriptsize\mbox{se}}_M\right)$, $\varrho$}
j=1\;
\While{M>0}{
$C^{\scriptsize\mbox{cl}}_j=C^{\scriptsize\mbox{se}}_1$\;
$n=0$\;
\For{$i=2,\ldots, M$ }{
  \If{$\mathcal{D}(C^{\scriptsize\mbox{se}}_{1},C^{\scriptsize\mbox{se}}_{i} )\leq \varrho$}{
  $n=n+1$\;
  $n_2=x^{\scriptsize\mbox{se}}_{2}-x^{\scriptsize\mbox{se}}_{1}$\;
  $n_1=x^{\scriptsize\mbox{se}}_{i+1}-x^{\scriptsize\mbox{se}}_{i}$\;
  $C^{\scriptsize\mbox{se}}_i\implies \left(s_1(z_1), \ldots, s_1(z_{n_1+n_2})\right)$\;
  $C^{\scriptsize\mbox{cl}}_j\implies \left(s_2(z_1), \ldots, s_2(z_{n_1+n_2})\right)$\;
  $C^{\scriptsize\mbox{cl}}_j=\sum_{h=1}^{n_1+n_2}(s_1(z_h)+s_1(z_h))/2$\;
  Delete $C^{\scriptsize\mbox{se}}_{i}$ from $\mathbf{C}^{\scriptsize\mbox{se}}$\;
  }
  }
  $j=j+1$.
}
\caption{Signal distance based clustering}
\label{native_clustering}
\end{algorithm}

\subsubsection{Results}
\begin{figure}[h]
\centering
\begin{subfigure}{0.4\columnwidth}
\centering
\includegraphics[width=\textwidth]{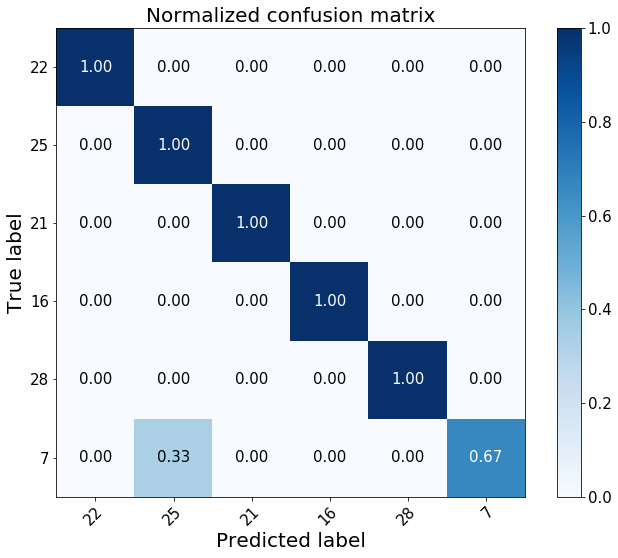}
\caption{}
\end{subfigure}
\begin{subfigure}{0.4\columnwidth}
\centering
\includegraphics[width=\textwidth]{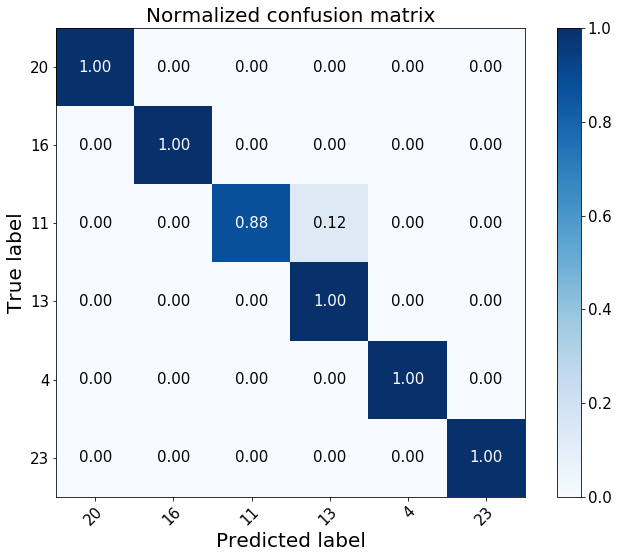}
\caption{}
\end{subfigure}
\begin{subfigure}{0.4\columnwidth}
\centering
\includegraphics[width=\textwidth]{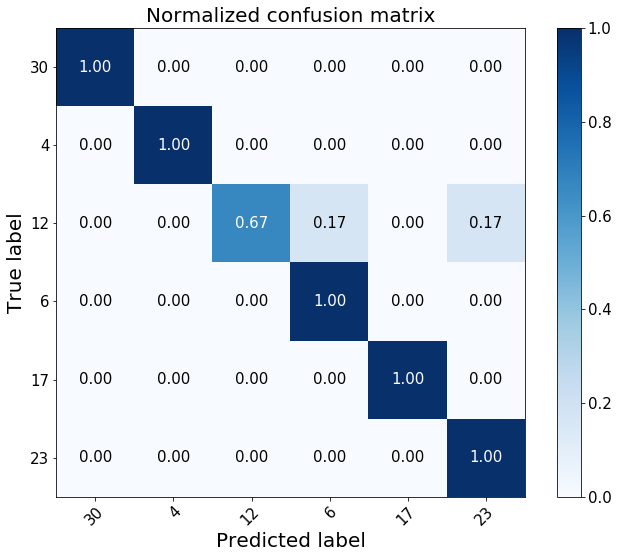}
\caption{}
\end{subfigure}
\begin{subfigure}{0.4\columnwidth}
\centering
\includegraphics[width=\textwidth]{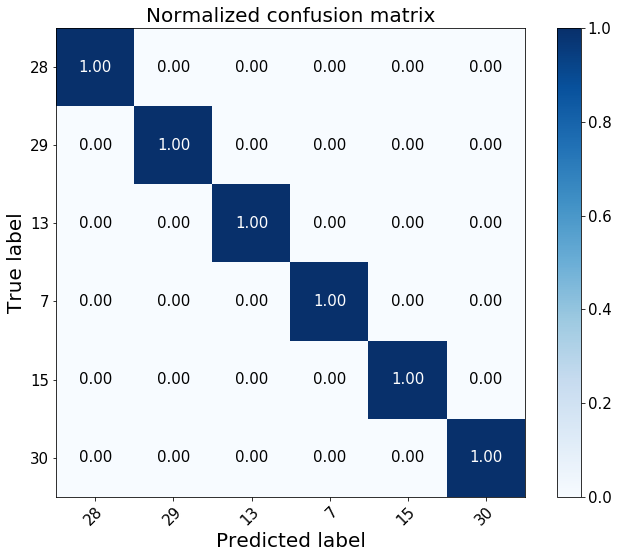}
\caption{}
\end{subfigure}
\caption{Confusion matrix for 6-class classification. The classification is applied on x-axis of accelerometer walking signal cut by  x-axis accelerometer based segmentation. Top left (a): volunteers involved are 22, 25, 21,16, 28, 7; accuracy and F1 score are 0.9714 and 0.9538 respectively. Top right (b): volunteers involved are 20, 16, 11, 13,  4, 23; accuracy and F1 score are 0.9750 and 0.9761 respectively. Bottom left (c): volunteers involved are 30,  4, 12,  6, 17, 23; accuracy and F1 score are 0.9500 and 0.9440 respectively. Bottom right (d): volunteers involved are 28, 29, 13,  7, 15, 30; accuracy and F1 score are 0.9500 and 0.944 respectively. Accuracy and F1 score are 1 and 1 respectively. }
\label{confusion_matrix_example}
\end{figure}

We discuss the results from several experiments under different scenarios in this subsection. Testing samples are single cycle based. Note that all the following experiments are conducted based on full-gait x-axis accelerometer segmentation, see Section \ref{segmentation} for details.   
\paragraph{Experiment 1: 6, 10, and 30 classes user-adversary identification (classification) using x-axis accelerometer and x-axis gyroscope respectively  for walking signal only.}Results are shown in Table \ref{result_table}. We run 6, 10, and 30 classes classification on x-axis accelerometer and x-axis gyroscope walking signals respectively to identify the user and the adversary simultaneously. The x-axis accelerometer or x-axis gyroscope walking signals is segmented by x-axis accelerometer based segmentation. As the number of classes or the number of volunteers involved in the experiments increases, more potential adversaries are considered and therefore, adding more complicity to the classification system. This can be verified by the decreased accuracy from around 0.97 to 0.9 as the number of class rises from 6 to 30. All the performance results in the table are the mean values and standard errors of the results from 20 times randomly pick 6, 10, or 30 out of 30 volunteers in HAPT data set. In Figure \ref{confusion_matrix_example}, several confusion matrices of the classification results for each of the 4 groups contains randomly picked 6 volunteers are demonstrated. 
Although, multi-level classification seems to be more complicated than the binary one, our technology still reserves very high accuracy. 

Besides the multi-level classification to identify user and adversary, we can also demonstrate a binary classification for user only identification; while we keep the training process the same, testing set includes the test cycles of the true user and randomly pick 2 test cycles of each of the 29 potential adversaries. The results are of 0.9744 and 0.9789 on average of test results for accelerometer and gyroscope signals respectively; the results perform better than that in 
\cite{neverova2016learning, damavsevivcius2016smartphone} which have accuracy of 0.9302 and 0.943 respectively.

\begin{table}[h!]
  \centering
  \begin{tabular}{ccccccccc}
    \toprule
     Number  &\multicolumn{4}{c}{Accelerometer}
     &\multicolumn{4}{c}{Gyroscope}   \\
    \cline{2-9}
    of class&  ACC & PPV & TPR & F$1$&ACC & PPV & TPR & F$1$ \\
    \midrule
    \multirow{2}{*}{6}&0.9649  &0.9730 &0.9558 &0.9592 &0.9706 & 0.9759 & 0.9620 &0.9647\\
    &(0.008) &(0.006)&(0.009)&(0.008)&(0.007)& (0.006)& (0.009)& (0.008)\\
    \multirow{2}{*}{10}&0.9348  &0.9490 &0.9247 &0.9270 &0.9394 & 0.9528 & 0.9284 &0.9326\\
    &(0.008) &(0.006)&(0.009)&(0.009)&(0.008)& (0.006)& (0.009)& (0.009)
     \\
    30& 0.8789& 0.8986& 0.8677& 0.8684 & 0.9000&0.9190&0.8835&0.8890   \\
    \bottomrule
  \end{tabular}
  \caption{Identification (classification) on different number of classes on x-axis accelerometer and x-axis gyroscope walking signal respectively. The signal type "accelerometer" means the identification is applied on x-axis of accelerometer signal cut by the x-axis accelerometer based segmentation; the signal type "gyroscope" means the identification is applied on x-axis of gyroscope signal cut by the x-axis accelerometer based segmentation. Each time we randomly pick 6, 10 or 30 out of the 30 volunteers to do the identification and  the performance measurements are mean values of the results of 20 times repeating of the process; the standard errors are given in parentheses. "ACC": accuracy; "PPV": precision; "TPR": recall; "F$_1$": harmonic mean of precision and recall.  }
   \label{result_table}
\end{table}

\paragraph{Experiment 2: 6, 10, 30 classes user-adversary identification (classification) using 3-axis accelerometer and 3-axis gyroscope respectively  for walking signal only.}Results are shown in Table \ref{result_table_mix_axis}. Similar to  Experiment 1, but use three axes of accelerometer or three axes of gyroscope signal in the classification process. Each test cycle (contains 3 axes as a group) is labeled to the volunteer that has the closet archetype (from 3 axes)  to at least one of the axes of the test cycle.  All the results are improved compared with the results using x-axis data only. The binary classification established the same as that in Experiment 1 provides an accuracy of 0.9906 for both accelerometer and gyroscope type classification.

\begin{table}[h!]
  \centering
  \begin{tabular}{ccccccccc}
    \toprule
     Number  &\multicolumn{4}{c}{Accelerometer}
     &\multicolumn{4}{c}{Gyroscope}   \\
    \cline{2-9}
    of class&  ACC & PPV & TPR & F$1$&ACC & PPV & TPR & F$1$ \\
    \midrule
    \multirow{2}{*}{6}&0.9812& 0.9838& 0.9833& 0.9811 &0.9879& 0.9900& 0.9889& 0.9880\\
    &(0.006) &(0.005)&(0.005)&(0.006)&(0.005)& (0.004)& (0.004)& (0.004)\\
    \multirow{2}{*}{10}&0.9798& 0.9848&  0.9783& 0.9785 &0.9767& 0.9822& 0.9775&     0.9766\\
    &(0.005) &(0.003)&(0.005)&(0.005)&(0.005)& (0.004)& (0.005)& (0.005)
     \\
    30& 0.9573& 0.9659& 0.9556& 0.9537& 0.9573& 0.9659& 0.9556& 0.9537  \\
    \bottomrule
  \end{tabular}
  \caption{Identification (classification) of different number on 3-axis accelerometer and 3-axis gyroscope walking signal respectively. The signal type "accelerometer" means the identification is applied on 3-axis accelerometer signal cut by the x-axis accelerometer based segmentation; the signal type "gyroscope" means the identification is applied on 3-axis of gyroscope signal cut by the x-axis accelerometer based segmentation. Each time we randomly pick 6, 10 or 30 out of the 30 volunteers to do the identification and  the performance measurements are mean values of the results of 20 times repeating of the process; the standard errors are given in parentheses. "ACC": accuracy; "PPV": precision; "TPR": recall; "F$_1$": harmonic mean of precision and recall.  }
   \label{result_table_mix_axis}
\end{table}

\paragraph{Experiment 3: 6 classes user-adversary identification (classification) using 3-axis accelerometer and 3-axis gyroscope respectively for walking up, walking down and mixed walking signals.}Results are shown in Table \ref{result_table_more_mix_axis}. The experiments on walking up and walking down signals are similar to Experiment 1 with 6 classes but replace the activity with walking up and walking down respectively. The mixed walking experiment using the iterative segmentation presented in Figure \ref{mix_seg}; the classification is applied on three axes of accelerometer or gyroscope signal together for mixed walking activities.

\begin{table}[h!]
  \centering
  \begin{tabular}{ccccccccc}
    \toprule
     Signal  &\multicolumn{4}{c}{Accelerometer}
     &\multicolumn{4}{c}{Gyroscope}   \\
    \cline{2-9}
    Type&  ACC & PPV & TPR & F$1$&ACC & PPV & TPR & F$1$ \\
    \midrule
    \multirow{2}{*}{Walking up}&0.9423 &0.9632& 0.9417& 0.9388 &0.9385& 0.9569& 0.9375 &    0.9336\\
    &(0.015) &(0.009)&(0.015)&(0.016)&(0.013)& (0.010)& (0.013)& (0.014)\\
    \multirow{2}{*}{Walking down}&0.8140& 0.7715& 0.7936& 0.7604&0.8408 &0.7768& 0.7992& 0.7692\\
    &(0.032) &(0.042)&(0.032)&(0.038)&(0.020)& (0.034)& (0.025)& (0.030)
     \\
   \multirow{2}{*}{Mixed Walking}& 0.7170& 0.7444& 0.6936& 0.6928& 0.7903&0.8163&0.7857&0.7834   \\
   &(0.025) &(0.025)&(0.025)&(0.024)&(0.014)& (0.013)& (0.014)& (0.013)
     \\
    \bottomrule
  \end{tabular}
  \caption{Identification of different activites by accelerometer based full-cycle segmentation of walking signal. The signal type "accelerometer" means the identification is applied on 3-axis accelerometer signal cut by the accelerometer based segmentation; the signal type "gyroscope" means the identification is applied on 3-axis of gyroscope signal cut by the accelerometer based segmentation. Each time we randomly pick 6 out of the 30 volunteers to do the identification and  the performance measurements are mean values of the results of 20 times repeating of the process; the standard errors are given in parentheses. "ACC": accuracy; "PPV": precision; "TPR": recall; "F$_1$": harmonic mean of precision and recall. }
   \label{result_table_more_mix_axis}
\end{table}

     

\section{Discussion and Conclusion}
\label{discussion}
The technology developed here would have great impact on cell phone user-adversary identification and cybersecurity. 
Comparing with existing segmentation methods such as zero crossing and Lomb-Scargle Periodogram, we cut at the critical points based on geometric features integrated with statistical and iterative methods. Our method is more robust and scalable. Moreover, the new segmentation method is applicable for different walking-type signals such as walking, walking upstairs, walking downstairs and mixed walking. Further research involves applying our proposed technology on more flexible walking signals such as walking with cell phone in hand, in pocket or in bag. The signal distance proposed in this paper provides a new approach of measuring the difference between signals which gives a possibility of translating the way human vision in distinguishing signals to  machine language. It can be broadly used in all kinds of signal processing analysis which indicates many possible directions of further researches. The user-adversary identification algorithm we proposed easily detects if certain walking behaviors are from the user or not. More precisely, it is possible to determine who is this imposter if potential imposters are provided. Comparing to the fact of tuning a large number of hyper-parameters in Neutral Network methods, our approaches only have four parameters with clear geometric meaning involved; the parameters can be tuned or given based on geometric analysis integrated by statistics. We obtain an accuracy of 0.9879 for 6-class classification on walking signal using 3 axes of the both accelerometer and gyroscope data; the accuracy is of 0.9812 when using only 3 axes of the accelerometer data; we still keep an accuracy of 0.9649 or 0.9706 when only using x-axis of the accelerometer or x-axis of accelerometer and gyroscope data. Even if the classification is on all the possible classes (30 classes) in HAPT, we still reserve an accuracy of 0.9573 by our approaches. When the multi-level classification degenerates to binary classification, we obtain an accuracy of around 0.99 for walking signal. It is worth to point out that the walking up and walking down data sets are very small, yet our experiments still showing solid results while there is not enough data for training in Neutral Network methods. In summary, the techniques we developed here is in high accuracy  in user-adversary identification even with multi-level classification, which is crucial in cybersecurity.

\bibliographystyle{unsrt}
\bibliography{bib}

\end{document}